\pdfoutput=1
\documentclass[prl,reprint,superscriptaddress,nobalancelastpage,showpacs]{revtex4-1}

\usepackage{color,amsthm,amsmath,amsfonts,graphicx,bm,amssymb}
\usepackage[plainpages=false]{hyperref}
\usepackage{epstopdf}
\usepackage[normalem]{ulem}
\usepackage{cancel}
\usepackage{comment}
\usepackage{etoolbox}
\usepackage{filecontents}
\usepackage{soul}
\usepackage{MnSymbol} 
\usepackage{mathtools}

\makeatletter
\newcommand{\specialcell}[1]{\ifmeasuring@#1\else\omit$\displaystyle#1$\ignorespaces\fi}
\makeatother

\newcommand{\beq}{\begin{equation}}
\newcommand{\eeq}{\end{equation}}

\def\Ai{\operatorname{Ai}}
\def\Prob{\operatorname{Prob}}
\def\KAi{K_{\mbox{\tiny Ai}}}
\def\rhotail{\rho}

\def\rhoGUE{\rhotail^{\mbox{\tiny GUE}}}
\def\rhoDP{\rhotail^{\mbox{\tiny DP}}}
\def\trhoGUE{{\tilde\rhotail}^{\mbox{\tiny GUE}}}
\def\trhoDP{{\tilde\rhotail}^{\mbox{\tiny DP}}}
\def\PD{P}
\def\PDGUE{\PD^{\mbox{\tiny GUE}}}
\def\PDDP{\PD^{\mbox{\tiny DP}}}
\def\tPDGUE{\tilde{\PD}^{\mbox{\tiny GUE}}}

\def\Det{\operatorname{Det}}

\def\pv{\boldsymbol\mu}
\def\lv{\boldsymbol\lambda}
\def\xv{\mathbf{x}}
\def\yv{\mathbf{y}}

\def\kv{\mathbf{k}}
\def\mv{\mathbf{m}}

\def\gv{\boldsymbol{\gamma}}

\def\ssum{\gamma}
\def\pp{p}

\def\cbar{\bar c}
\def\partN{n}
\def\num{N}

\def\partNS{n_s}

\def\mom{m}

\def\Tr{\operatorname{Tr}}

\def\matrN{\mathcal{N}}

\def\sym{\operatorname{\sigma}}

\def\pp{p}
\def\dis{\eta}

\newcommand{\be}{\begin{equation}}
\newcommand{\ee}{\end{equation}}
\newcommand{\bea}{\begin{eqnarray}}
\newcommand{\eea}{\end{eqnarray}}

\newcommand{\titleinfo}{Mutually avoiding paths in random media and largests eigenvalues 
of random matrices}
\graphicspath{{../},{figures/}}
\begin{document}

\title{\titleinfo} 

\author{Andrea De Luca}
\email{andrea.deluca@lptms.u-psud.fr}
\affiliation{Laboratoire de Physique Th\' eorique et Mod\` eles Statistiques (UMR CNRS 8626), Universit\' e Paris-Sud, Orsay, France}

\author{Pierre Le Doussal}
\email{ledou@lpt.ens.fr}
\affiliation{Laboratoire de Physique Th\'eorique de l'ENS, CNRS \& Ecole Normale Sup\'erieure de Paris, Paris, France.}

\date{\today}

\begin{abstract}	
Recently, it was shown that the probability distribution function (PDF) of the free energy of a single continuum directed polymer (DP) in a random potential, equivalently of the height of a growing interface described by the 
Kardar-Parisi-Zhang (KPZ) equation, converges at large scale to the Tracy-Widom distribution.
The latter describes the fluctuations of the largest eigenvalue of a random matrice, drawn from
the Gaussian Unitary Ensemble (GUE), and the result holds for a DP with fixed endpoints, 
i.e. for the KPZ equation with droplet initial conditions. 
A more general conjecture can be put forward, relating the free energies of $N>1$ non-crossing continuum DP
in a random potential, to the $N$-th largest eigenvalues of the GUE. Here, using replica methods,
we provide an important test of this conjecture 
by calculating exactly the right tails of both PDF's and showing that they 
coincide for
arbitrary $N$.
\end{abstract}


\maketitle

\paragraph{Introduction. --- }

Remarkable connections have emerged in the last decade between random matrix theory, 
growth models, and glassy systems. 
The celebrated Kardar-Parisi-Zhang (KPZ) equation~\cite{KPZ}
provides the simplest description for the growth of an interface in presence of noise. This equation 
sits at the center of a wide universality class~\cite{kpzreviews}, encompassing 
several models and physical systems, such as the polynuclear growth model (PNG)~\cite{PNG}, 
the asymmetric exclusion processes (ASEP)~\cite{DerridaASEP1998,spohnStatTASEP,prolhac2009cumulants} and Burgers turbulence~\cite{Burgers}.

Additionally, the height $h(x,t)$ of the KPZ interface in $d$ dimensions can be exactly mapped into 
(minus) the free energy of a directed polymer (DP) of length $t$ in a quenched random potential in $1+d$ dimension ~\cite{exponent,directedpoly}.
The DP is one of the most straightforward realization of a glass, with applications
including domain walls in magnets~\cite{lemerle}, vortex lines in superconductors~\cite{BlatterVortexReview94}, 
localization paths in Anderson insulators~\cite{SoOr07} and even to problems in biophysics~\cite{hwasimilarity,*OtwinowskiKrug2014} and economics~\cite{GueudreExploit2014}.

The link between KPZ and DP has been particularly fruitful in $d=1$,
where a hidden integrable structure comes to light. In this case, several exact solutions, 
first for zero temperature ~\cite{Johansson2000}, and later 
for finite temperature discrete ~\cite{logsep2,logboro,usLogGamma} and continuum
DP models ~\cite{we,dotsenko,corwinDP,we-flat,SasamotoStationary,reviewCorwin}, 
unveiled an astounding connection:
the probability distribution of the (scaled) KPZ height field $h(x,t)$ coincides with the 
(scaled) distribution of the largest eigenvalue
of a random matrix drawn from the famous Gaussian ensembles:
this is the so-called Tracy-Widom (TW) distribution \cite{TW1994}, recurring in broad variety of contexts~\cite{othersTW}.
In particular, if we define as $\hat {\cal Z}_1(t)$, the partition function in the continuum of a DP
which starts and end at the same point $x(t) = x(0)$ (see below for an explicit definition), 
it was found in \cite{we,dotsenko} that at large time one can write 
$\ln \hat {\cal Z}_1(t) \simeq - t/12 +  \hat \gamma_1 t^{-1/3}$,
where $\hat \gamma_1$ follows the $\beta=2$ TW distribution 
associated to the Gaussian Unitary Ensemble (GUE).
Here we use hat to denote a random variable. 
While these results have been confirmed by rigorous mathematical treatments
\cite{corwinDP,BorodinMacdo,Quastelflat}, 
a complete understanding of the deep reason behind this correspondence 
remains open. A natural direction to shed some light on this problem, 
is to extend this connection
beyond the maximal eigenvalue $\hat \gamma_1$, to the full portion 
of the GUE spectrum around the edge. 

In the present work, we consider an ensemble of $\num$ mutually-avoiding polymers, i.e.
several directed paths constrained not to intersect one another, and 
competing to optimize their total energy in the same random media.
We extend the study of the single polymer partition function $\hat {\mathcal{Z}}_1$, 
to the one of $N$ non-crossing paths $\hat{\mathcal{Z}}_N$. We build on a 
general method which we recently developed to treat 
any number $\num$ of DPs, but until now, only applied in the specific case $N=2$
to analyze the non-crossing probability \cite{UsNonCrossing1,supplmat_deluca2015,UsNonCrossing2}.

\begin{figure}[	t]
  \includegraphics[trim={0.5cm 0.9cm 0.2cm 0.2cm},clip,width=0.85\columnwidth]{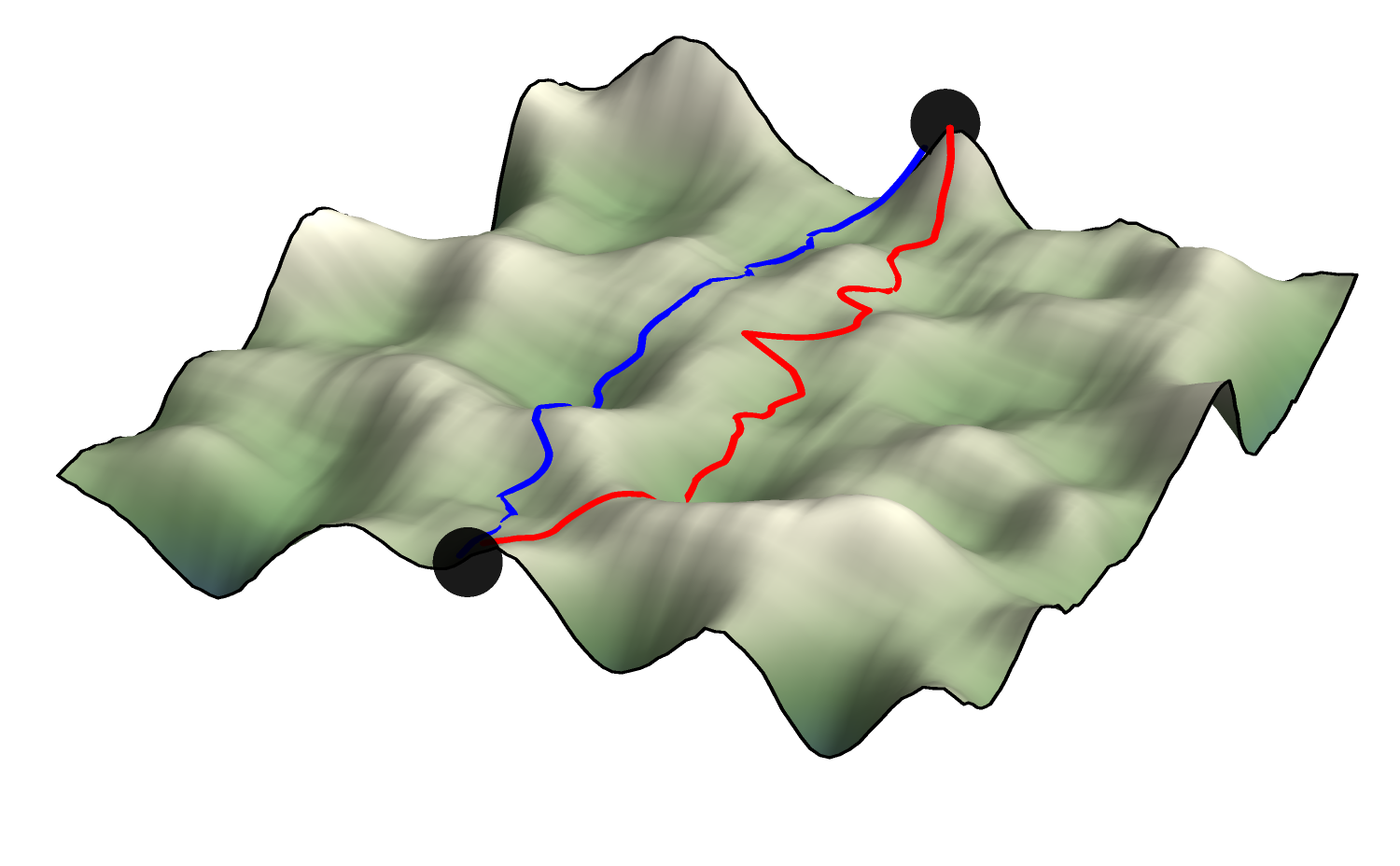}
  \caption{   (Color online)
  Sketch of the $N=2$ case: two polymers start and end at (almost) coinciding points. 
  The blue/red lines represent the point with maximal probability at any time slice $t$ and the 
  disorder realization is chosen to have well-separated paths. 
  }
  \label{fig:sketch}
\end{figure}

Here we put forward the conjecture that the $N$-path free energy takes the form at large time
\cite{footnote0}
\be \label{larget} 
\ln \hat {\cal Z}_\num(t) \simeq - \num t/12 + t^{1/3} \hat \zeta^{(\num)}
\ee
where the random variable $\hat \zeta^{(\num)}$ coincides in law with \textit{partial sum} of the $\num$ largest eigenvalues
$\hat \gamma_1,\ldots, \hat \gamma_N$ of a GUE random matrix 
\bea
\label{conjsum}
\hat \zeta^{(\num)} 
\stackrel{\mbox{\tiny in law}} { \equiv}\sum_{i=1}^\num \hat \gamma_i =: \hat \gamma 
\eea
The validity of this conjecture for the continuum, finite temperature model, is suggested by an argument of
universality \cite{OConnellWarrenMultiLayer,UsNonCrossing1,supplmat_deluca2015}, together with exact 
results on
discrete DP models {\it at zero temperature}, specifically the last passage
percolation model \cite{BaikJohansson2000,JohanssonPlancherel2001}
and the semi-discrete directed
polymer~\cite{OConnelYor2002,Doumerc,Benaych-Georges,logsep2}.

Obviously showing the equality of the probability distribution functions (PDF) $P^{DP}_N(\zeta)$ and $P^{GUE}_N(\gamma)$  
is a major challenge. Here we will provide a first test, by showing that
their leading (stretched exponential order) {\it tail} approximant functions are identical. More
precisely we will show that at large arguments 
\bea \label{tail1} 
&& \PDDP_\num(\zeta) = \rhoDP_\num(\zeta) (1 + O(e^{- a_N \zeta^{3/2}}) ) \\
&& \PDGUE_\num(\ssum) = \rhoGUE_\num(\ssum) (1 + O(e^{- a'_N \gamma^{3/2}}) ) \label{tail2}
\eea 
with $a_{\num},a'_{\num}>0$ and exactly the {\it same} function 
$\rhotail_\num^{\mbox{\tiny DP}}(\ssum)  = \rhotail_\num^{\mbox{\tiny GUE}}(\ssum) = O(e^{-\frac{4 \ssum^{3/2}}{3 \sqrt{N}}})$. 
Here and below
$O(e^{- a \gamma^{3/2}})$ means at leading exponential accuracy. 
Note that the function $\rhotail^{\mbox{\tiny GUE}}_{\num}(\ssum)$ is non-trivial,
hence the coincidence is a strong hint for the conjecture to
hold. For instance in the simpler case of $N=1$, where
the conjecture is known to hold, one has 
$\rhotail^{\mbox{\tiny GUE}}_{N=1}(\ssum)=\Ai'(\ssum)^2 - \ssum \Ai(\ssum)^2$. 
Likewise, we will provide (more complicated) formula for $N>1$. 

Note that non-intersecting Brownian motions (sometimes dubbed ``watermelon configurations'') have already 
been put in relation with Airy processes and Tracy-Widom distributions~\cite{schehr2008exact, *forrester2011non, *schehr2012extremes}. These studies hold however in a very different context, 
in particular in the absence of any quenched disorder.

\begin{figure}[t]
  \includegraphics[width=0.95\columnwidth]{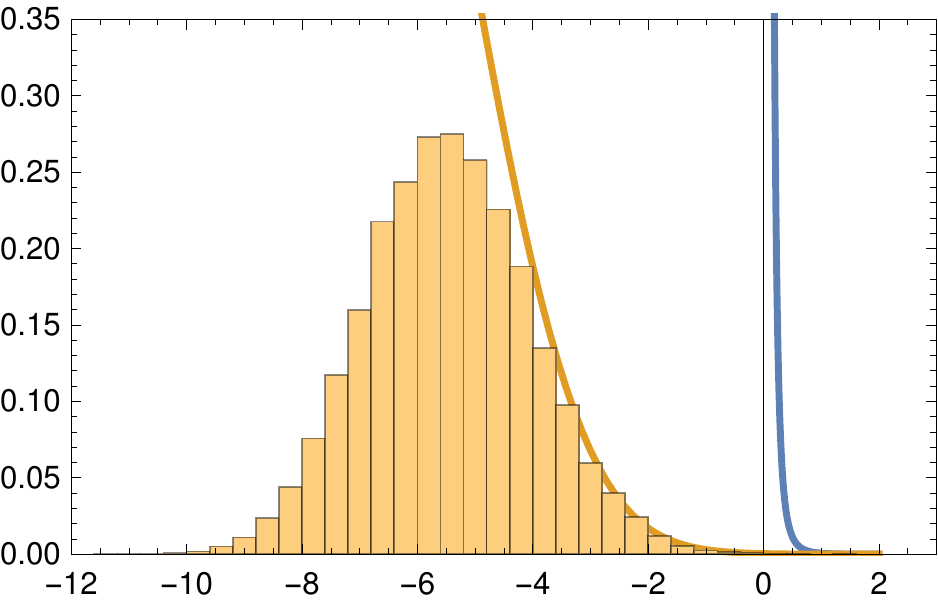}
  \caption{   (Color online) 
  The empirical distribution for the sum of the first $\num=2$ eigenvalues in $10^5$ realizations of
  GUE matrices of size $\mathcal{N} = 250$. The continuous line 
  are two different approximations for the tail obtained:
  from the inverse Laplace transformation of Eq.\eqref{hatp} (orange); from the simple approximation
  $R_N(\gamma)=1$ in 
  Eq.~\eqref{largesum} (blue).
  }
  \label{fig:comparison}
\end{figure}

\paragraph{The GUE ensemble. ---}
To fix the notation we take the GUE specified by the measure $\propto \exp(-\Tr H^2)(dH)$,
where $H$ is a complex $\matrN \times \matrN$ hermitian matrix.
For large $\matrN$, the support of the spectrum concentrates in $(-\sqrt{2\matrN}, \sqrt{2\matrN})$. Nevertheless,
there is a finite probability for the eigenvalues $\hat\lambda_1 > \ldots > \hat\lambda_\matrN$ 
to fall outside this interval. In particular,
introducing the rescaled eigenvalues 
$\hat\gamma_l = (\hat\lambda_l - \sqrt{2\matrN}) \sqrt{2} \matrN^{1/6}$, the mean spacing for the variables 
close to $\hat \gamma_1$ becomes of order unity. In the limit $\matrN \to \infty$, this 
results in a well-defined determinantal point process, characterized by the correlation functions
\cite{JohanssonReview,BorodinDet}
\begin{equation}
\label{rkdef}
r_\num(x_1,\ldots,x_\num) = \det[\KAi(x_i, x_j)]_{i,j=1}^\num \;.
\end{equation}
for the density probability that there is a scaled eigenvalue in 
each interval $[x_i,x_i+dx_i]$, $i=1,..N$. 
Here, the Airy Kernel has been introduced as
\begin{equation}
 \label{airykern}
 \KAi(x,y) = \int_0^\infty dw \Ai(x + w) \Ai(y + w)\;.
\end{equation}
From this expression, a simple application of the inclusion-exclusion principle 
permits expressing the joint probability distribution for the $\num$-largest (rescaled)
eigenvalues~\cite{witte2013joint} (non-vanishing and normalized to unity in the domain $\gamma_1>\ldots>\gamma_\num$)
\begin{equation}
\label{pkdef}
 \pp_\num(\gv_\num) = \sum_{k=0}^\infty \frac{(-1)^k}{k!} \prod_{j=1}^k \int_{\min_{i=1}^N \gamma_i}^\infty dx_j \;
 r_{\num+k}(\gv_\num, \xv_k) 
\end{equation}
where the bold symbol $\gv_\num $ stands for $\gamma_1, \ldots, \gamma_\num$ (and similarly for $\xv_p$). 
In the particular case $\num = 1$, this expression can be recast as the derivative of
a Fredholm determinant: $\pp_1(\gamma) \equiv f_2(\gamma)$ is the GUE Tracy-Widom function.
Setting $f_2(\gamma) = dF_2(\gamma)/d\gamma$, the cumulative distribution function $F_2(\gamma)$ is expressed as
\begin{equation}
F_2(\gamma) = \Det( 1 - \Pi_\gamma \KAi \Pi_\gamma) \;.
\end{equation}
with $\Pi_\gamma$ the projector onto $[\gamma, +\infty)$.
\paragraph{Sums of largest eigenvalues. ---}
We now introduce the partial sum of the $N$-largest eigenvalues, defined as
\begin{equation}
 \label{partialsum}
 \hat\ssum^{(\num)} = \hat\gamma_1 + \ldots + \hat\gamma_\num
\end{equation}
and in the following we will omit the superscript $\num$ when not explicitly necessary. 
The probability distribution $\PDGUE_\num(\ssum)$ for this quantity can be inferred from Eq.~\eqref{pkdef}
\begin{equation}
\label{sumdistr}
 \PDGUE_\num(\ssum) = \frac{1}{\num!}\int d\gamma_1\ldots \, d\gamma_N \,\delta\bigl(\ssum - \sum_{k=1}^\num 
 \gamma_k\bigr) \pp_\num(\gv_\num) \;.
\end{equation}
It is useful to introduce the double-sided Laplace Transform (LT) of $\PD_\num(\ssum)$ as
\begin{equation}
\label{lapltrans}
 \tPDGUE_{\num} (u) := \overline{\exp(u \hat \gamma)}= \int_{-\infty}^\infty d \ssum  \, \PDGUE_\num(\ssum) e^{u \ssum} \;.
\end{equation}
We are interested in the right-tail $\ssum \gg 1$, which governs the integral when $u$ is large. 
Because of the behavior of the tail 
$\KAi(\gamma_i,\gamma_j) \sim e^{-\frac{2}{3}( \gamma_i^{3/2} + \gamma_j^{3/2}) }$,
this regime is dominated by the configuration minimizing the sum $\sum_i \gamma_i^{3/2}$
at fixed $\gamma=\sum_i \gamma_i$:
this suggests that large values of the sum $\gamma$ require 
all the $N$ largest eigenvalues to be of the same order of magnitude, i.e.
$\gamma_k \simeq \ssum/\num$.
Then, in order to estimate the tail $\rhoGUE_\num(\ssum)$ defined by (\ref{tail1}) of the distribution of the sum in \eqref{sumdistr}, 
we can limit the expansion in \eqref{pkdef} to the first term $\pp_\num(\gv_\num) \simeq r_\num(\gv_\num)$.
%
Rearranging the determinant in $r_\num(\gv_\num)$, we obtain
\bea
\label{hatp}
 && \trhoGUE_\num (u) =\\
&& \frac{1}{N!} \prod_{i=1}^\num  \int_0^\infty dv_i 
 \det \big[ \int_{-\infty}^\infty d\gamma e^{\gamma u} \Ai(\gamma + v_j) \Ai(\gamma + v_k) \big]_{j,k=1}^\num 
 \nonumber
 \eea
 which, after some simple manipulations \cite{SuppMat} leads
 to our main result
 \be \label{main}
\trhoGUE_\num (u) = \frac{e^{\frac{\num u^3}{12}}u^{-\frac{3N}{2}}}{\pi^{\num/2} \num!}
 \prod_{i=1}^\num \int_{v_i>0} \; e^{-2 v_i} \det \left[
 e^{-\frac{(v_j-v_k)^2}{u^3}} \right]_{j,k=1}^\num
\ee
which generalizes the remarkably simple $N=1$ result for the LT of the tail
of the Tracy-Widom distribution
\bea
\trhoGUE_{N=1} (u) = \frac{e^{\frac{u^3}{12}}}{2 \pi^{1/2} u^{\frac{3}{2}}}
\eea 
For $N>1$ it takes the general form
\begin{equation}
\label{pNtilde1}
 \trhoGUE_N (u) = \frac{e^{\frac{N u^3}{12}} G(N+1)}{2^{N(N+1)/2} \pi^{N/2} u^{3N^2/2}}
 Q_N\Bigl(\frac{1}{u^3}\Bigr) 
 \end{equation}
where $Q_N(0)=1$ and $Q_N(z)$ admits a series expansion around $z=0$,
since at large large $u$, this last determinant can be computed explicitly \cite{SuppMat}.
The Laplace inversion of (\ref{pNtilde1}) gives the general form of the tail function
$\rhoGUE_\num(\ssum)$ where the leading behavior at large $\gamma$ is apparent 
(with $R_N(+\infty)=1$)
\begin{equation}
\label{largesum}
\rhoGUE_\num(\ssum) = \frac{ N^{\frac{3 N^2-1}{4}} G(N+1) 
e^{-\frac{4 \ssum^{3/2}}{3 \sqrt{N}}}}{  2^{2\num^2} (2\pi)^{\frac{N+1}{2}}\ssum^{\frac{3 N^2+1}{4}}}
R_N(\gamma)
\end{equation}
where $G(x)$ is the Barnes function. The function $R_N(\gamma)$ can be obtained 
from subdominant orders in a saddle point expansion and has the form of
a double series in $1/\gamma$ and $1/\gamma^{3/2}$. 
In Fig.~\ref{fig:comparison}, we compare these predictions with the
empirical distribution for $N=2$. Note that the exact form
of $\rhoGUE_\num(\ssum)$ is a major improvement compared
to the naive approximation for the tail obtained by setting $R_N(\gamma)=1$ in (\ref{largesum}). 
From considerations of Airy function asymptotics it is easy to see that the corrections in (\ref{tail1}) to $\rhoGUE_\num(\ssum)$ itself, calculated 
as above, are indeed subdominant by $O(e^{- a'_N \gamma^{3/2}})$ with $a'_N=\frac{2}{3} N^{-3/2}$.

\paragraph{Mutually avoiding directed polymers. ---}
We introduce the partition function of a directed polymer with fixed endpoints $x, y$
\begin{equation}
 \label{Zorig}
 \hat Z_\dis(x;y|t) \equiv \int_{x(0) = x}^{x(t) = y} Dx e^{-\int_0^t d\tau \left[ \frac{1}{4}(\frac{dx}{d\tau})^2 - \sqrt{2 \cbar} \hat\dis(x(\tau), \tau)\right]} 
 \end{equation}
in a given realization of a random potential with white-noise correlations $\overline{\hat\dis(x,t) \hat\dis(x',t')} = \delta(x-x')\delta(t-t')$.  
In the following, to simplify the notation we rescale time and space and set $\cbar = 1$.
Considering $N$ polymers starting respectively at $\xv = x_1,\ldots, x_N$ and arriving at $\yv = y_1,\ldots, y_N$
the partition function constrained to non-intersecting paths can be expressed, using \cite{KarlinMcGregor}, as a single determinant
\begin{equation} \label{ZN} 
 \hat Z_\eta^{(\num)}(\xv; \yv| t) = \det [ \hat Z_\eta(x_i; y_j| t) ]_{i,j=1}^N \;.
\end{equation}
This expression involves arbitrary space dependence; in order to simplify it, we consider therefore the limit where
all the initial/final points coincide: $x_i = y_i = \epsilon u_i$. In the limit $\epsilon \to 0$, 
$ \hat Z_\eta^{(\num)}(\xv; \yv| t)  \simeq \frac{\epsilon^{N(N-1)}}{G(N+1)^2} \prod_{i<j} (u_i - u_j)^2 
\hat{\mathcal{Z}}_\num(t)$, where \cite{OConnellWarrenMultiLayer,UsNonCrossing1,supplmat_deluca2015,warren}
\begin{equation} \label{det1} 
 \hat{\mathcal{Z}}_\num(t) = \det[\partial_x^{i-1} \partial_y^{j-1} \hat Z_\eta(x;y|t)|_{x=y=0}]_{i,j=1}^N
 \;.
\end{equation}
This random variable will be our quantity of interest. 
Its integer moments can be treated 
in the framework of the nested Bethe ansatz (NBA) \cite{UsNonCrossing1}
and of Macdonald processes \cite{BorodinMacdo}.
As showed in \cite{UsNonCrossing1},
both methods lead to an expansion in terms of a sum over eigenstates of the (integrable) quantum Hamiltonian
associated to the attractive $\delta$-Bose gas, i.e. the Lieb-Liniger model. In particular, 
using a residue expansion of the contour-integral formula of ~\cite{BorodinMacdoRemark},
one obtain a series over integer partitions:
\begin{multline}  \label{partsum} 
\overline{\hat{\mathcal{Z}}_\num(t)^\mom} = \sum_{n_s=1}^n \frac{n!}{n_s! (2\pi )^{n_s}} \sum_{(m_1,\dots m_{n_s})_n} 
\\ 
 \prod_{j=1}^{n_s} \int_{-\infty}^{+\infty} \frac{d k_j}{m_j}e^{-t E[\kv, \mv]}
 \Phi[\kv, \mv]\, \mathcal{B}_{\num,\mom}[\kv, \mv]  
\,, 
\end{multline} 
where $(m_1, \ldots ,m_{n_s})_n$, indicates sum over all 
integers $m_j \geq 1$ whose sum equals $\sum_{j=1}^{n_s} m_j = \partN = \mom \num$ and the energy of the string configuration has the form
$E[\kv, \mv] =
\sum_{j=1}^{n_s} m_j k_j^2 + \frac{1}{12} (m_j - m_j^3)$.
Eq.~\eqref{partsum} can be interpreted as an expansion over Lieb-Liniger eigenstates composed by $n_s$ strings of sizes $m_1,\ldots, m_{n_s}$.
Then, the factor $\Phi[\kv, \mv]$ can be obtained from the normalization of the string eigenstates
and has the form \cite{we,cc-07}
\begin{equation}
\label{Phidef}
\Phi[\kv,\mv]=
\prod_{\substack{1\leq i<j\leq n_s}} 
\frac{(k_i-k_j)^2 +(m_i-m_j)^2 /4}{(k_i-k_j)^2 +(m_i+m_j)^2 /4} \;.
\end{equation}
The factor $\mathcal{B}_{\num,\mom}[\kv, \mv]$ 
encodes the non-crossing constraint and contains all the dependence on $\num$ and $\mom$.
It is expressed by introducing ($\mu_{jk} = \mu_j - \mu_k$)
\begin{equation}
\label{Bdef}
\mathcal{B}_{\num,\mom}[\pv] = \frac{1}{\num!^\mom} \sym_{\pv} \left[ 
\frac{\prod_{i=1}^\mom\prod_{(i-1)\num < j < k\leq i \num} h(\mu_{jk})}
{\prod_{1 \leq j < k\leq n} f(\mu_{kj})}\right]
\end{equation}
where the functions $h(u) = u(u+i)$ and $f(u) = u/(u+i)$ and 
$\sym_{\lv}[W(\lv)] = \sum_{R} W(R\lv)/\partN!$ is the symmetrization of $W(\lv)$ 
over the variables $\lv$. Then, to obtain $\mathcal{B}_{\num,\mom}[\kv, \mv]$,
one needs to specialize the $\partN$ variables $\pv = \{\mu_1,\ldots, \mu_\partN\}$ with 
$\{\tilde k_1, \tilde k_1 + i,\ldots, \tilde k_1 + i(m_1-1), \tilde k_2, \tilde k_2 + i,\ldots\}$ and
$\tilde k_j = k_j - i(m_j-1)/2$.

The standard way to extract the PDF of the random variable $\hat \zeta$
in Eq. (\ref{larget}) from the knowledge of the moments in Eq.~\eqref{partsum}
is to introduce a generating function by
\begin{equation}
 \label{genfun}
 g_\num(s) = \sum_{\mom = 0}^\infty \frac{(-1)^\mom}{m!} x^m \overline{\hat{\mathcal{Z}}_\num(t)^\mom }
 =\overline{\exp\bigl(-e^{- \lambda s +  t^{1/3}\hat \zeta}\bigr)}
\end{equation}
where $x$ is related to $s$ by 
$ x e^{-\frac{N t}{12}} = e^{-\lambda s} $ and we introduce the rescaled time $\lambda = (N t/4)^{1/3}$. 
Eq.~\eqref{genfun} has two advantages: i) it lifts the constraint over the sum of $m_i$ 
in Eq.~\eqref{partsum}; ii) in the limit $t\to\infty$,
$g_\num(s) \to \Prob(\hat\zeta < (\num/4)^{1/3} s)$, i.e. the cumulative distribution function (CDF)
of the random variable $\hat \zeta$. Unfortunately, even without the constraint, it is difficult to perform the sum
(\ref{genfun}) exactly for $N>1$: indeed, already obtaining a closed expression for $\mathcal{B}_{\num,\mom}[\pv] $
is a non-trivial task, which, apart from $\mathcal{B}_{1,\mom}[\pv] = 1$, 
has been overcome only for $\num=2$ \cite{UsNonCrossing2}, 
where however the sum in Eq.~\eqref{partsum} remains an open challenge.

Fortunately however, we can still deal with (\ref{genfun}) by replacing 
each moment with its asymptotics at large time. Although this {\it does not}
give the exact large time behavior of $g_N(s)$, it
is sufficient, as discussed below, to obtain the {\it exact tail behavior} of the PDF of $\hat \zeta$. 
Indeed, such properties already appear in the studies of the case $N=1$, 

At large times $t$ and for fixed $N,m$, the sum in \eqref{partsum} is dominated by
the configurations $\mv$ with smallest energy $E[\kv, \mv]$. 
In general, the energy $E[\kv, \mv]$ will be minimized
by the configurations with the largest possible $m_i$.
For a single polymer $\num = 1$, this simply translates into $\partNS = 1$ and $m_1 = \partN$.
However, for $\num > 1$, this configuration gives a vanishing contribution: 
a general property of $\mathcal{B}_{\num,\mom}[\kv, \mv]$
is that it vanishes on any configuration with at least one $m_j>m$ \cite{UsNonCrossing2}.
This condition has the simple physical interpretation: a bound state (i.e. a string) cannot 
be formed joining particles which have been constrained to avoid each other.
Surprisingly, this property is sufficient to completely 
determine the value of $\mathcal{B}_{\num,\mom}[\kv, \mv]$ on the 
lowest energy configuration with non-vanishing contribution, which is the one consisting of a set of 
$N$ $\mom$-strings, i.e. $\partNS = \num$ and every $m_j = \mom$ \cite{SuppMat}. 
Combining Eq.~(\ref{Phidef}) with  Eq.~(\ref{Bdef}) on this configuration, we have (omitting now the trivial dependence on $m_j= \mom$, and noting $k_{ij}=k_i-k_j$)
\begin{equation} \label{resB2} 
\Phi[\kv]\, \mathcal{B}_{N,\mom}[\kv] = 
\frac{\mom!^N}{(\mom N)!}\prod_{1 < i<j < \num} (-\imath k_{ij})_m (\imath k_{ij})_m
\end{equation}
where $(x)_m$ indicates the Pochhammer symbol. 
Inserting in the formula for the $m$-th moment (\ref{partsum}) and 
keeping only the configuration $m_j=m$, $j=1,..n_s$ 
with $n_s=N$, one finds $\mathcal{Z}^{(0)}_{N,m}(t)$, defined as the
leading contribution at large t and fixed $N,m$, to $\overline{ \hat{\mathcal{Z}}_{\num} (t)^\mom}$
in Eq.~\eqref{partsum} (see \cite{SuppMat}).

We now calculate $g^0_\num(s) = \sum_{\mom = 0}^\infty (-1)^\mom x^m
\mathcal{Z}^{(0)}_{N,m}(t)$. In order to deal with the summation over $m$ 
we follow two steps: i) 
we use the Airy trick \cite{we,dotsenko} to get rid of the factor $\mom^3$ in the exponent:
\begin{equation}
\label{airytrick}
\int_{-\infty}^\infty dy \Ai(y) e^{y w} = e^{w^3 /3} \;;
\end{equation}
ii) we rewrite the sum over $\mom$ using the Mellin-Barnes representation
\begin{equation}
 \label{MBtrick} 
 \sum_{\mom\geq1} (-1)^\mom {\sf f}(\mom) = 
 -\frac{1}{2i}\int_{\epsilon - i \infty}^{\epsilon + i \infty} \frac{dz}{\sin(\pi z)} {\sf f}(z)\;.
\end{equation}
where $\epsilon \in ]0,1[$ has to be chosen such that 
the function ${\sf f}(z)$ does not have singularities for $\Re(z)>\epsilon$.
After some manipulations (see \cite{SuppMat}), one arrives at
\begin{multline}
\label{gnexpr}
 g^{(0)}_N(s) \stackrel{t \to \infty}{=}
 1- \frac{1}{\num!}\prod_{i=1}^{\num} \int_{-\infty}^{+\infty} \frac{dk_i}{2\pi} 
   \int_{0}^\infty dy \Ai\bigl(y+\sum_i k_i^2 + s\bigr) \\
   \int_{\epsilon - \imath \infty}^{\epsilon + \imath \infty} \frac{dz}{2\pi \imath z} 
e^{\sqrt{N} z y} \det \left[\frac{1}{2z + \imath k_{jk}}\right]_{j,k=1}^\num \;.
\end{multline}
We observe how a nice determinantal structure emerges at this level, reminiscent of
the $\num\times\num$ determinant appearing in Eq.\eqref{hatp}. To compare further,
we obtain the PDF by differentiating with respect to $s$ and
we take again the Laplace transform 
\begin{equation}
\label{lapltransDP}
\trhoDP_\num(u) = \overline{\exp(u \hat \zeta)} = \int_{-\infty}^\infty ds \; \partial_s g^{(0)}_\num(s) 
e^{(\frac{N}{4})^{1/3} u s} \;.
\end{equation}
The integral over $s$  in Eq.~\eqref{lapltransDP} 
can now be computed by a simple variation of Eq.~\eqref{airytrick}:
\begin{equation}
\label{aitrickdersimple}
  \int_{-\infty}^\infty ds \Ai'(y+ \sum_i k_i^2+s) e^{\tilde u s} =  
   - \tilde u e^{-\tilde u (\sum_i k_i^2 + y)} 
  e^{\frac{{\tilde u}^3}{3}}\;, 
\end{equation}
where in order to simplify the notation we set $\tilde u \equiv (N/4)^{1/3} u$.
When inserting this equality back in Eq.~\eqref{lapltransDP},
the integral over $y$ can be easily performed as $\epsilon>0$ and leads to a simple pole in 
at $z = \tilde{u}/\sqrt{N}$. This allows us to perform the integral over $z$, by closing the contour 
in the positive $\Re[z]$ half-plane and arrive at
\begin{equation}
\trhoDP_\num(u) = 
\frac{e^{\frac{\tilde{u}^3}{3}}}{\num!} \prod_{i=1}^{\num} \int_{-\infty}^{+\infty} \frac{dk_i}{2\pi} e^{-\frac{\tilde u k_i^2}{\num}} 
\det \left[\frac{1}{2 \tilde u + \imath k_{jk}}\right]_{j,k=1}^\num 
\end{equation}
We now check that this expression is equivalent to Eq.~\eqref{main}. Indeed,
expanding the determinant in a sum over the permutation group $\mathcal{S}_\num$ 
of $N$ elements through the Leibniz formula 
and introducing auxiliary variables $v_1,\ldots, v_\num$, we have
\begin{equation}
 \label{detnew}
\det \left[\frac{1}{2 \tilde u + \imath k_{jk}}\right]
= \sum_{P\in\mathcal{S}_\num} (-1)^{\sigma_P}
\prod_{j=1}^\num \int_0^\infty e^{-2\tilde u v_j -\imath( k_{j} - k_{ P_j})v_j}
\end{equation}
where $\sigma_P$ is the siganture of $P$.
We can now easily perform the gaussian integrals over the $k_1,\ldots, k_\num$ variables
and, relabeling $P \to P^{-1}$ in the sum, 
one obtains exactly the expansion of the determinant
in Eq.~\eqref{main} (see \cite{SuppMat} for more details), i.e.
the two Laplace transforms coincide $\trhoGUE_\num (u) =  \trhoDP_\num (u)$..
Via a Laplace inversion, this shows our main statement,
below Eq.~\eqref{conjsum}, namely that the two
PDF exactly coincide in the tails, i.e. 
$\rhoGUE_\num (\gamma) = \rhoDP_\num (\gamma)$.

Note that we have assumed that the restriction 
to the $N$ $m$-string states, gives the exact tail of the PDF of $\hat \zeta$
at large time, in other words that
$\lim_{t \to +\infty} g^{(0)}_N(s)|_{s=(4/N)^{1/3} \zeta} = 
1- \int_{\zeta}^{+\infty} d\zeta' \rhoDP_\num (\zeta')$,
and that the neglected terms give a contribution 
subdominant by $O(e^{- a_N \zeta^{3/2}})$ as in
(\ref{tail2}). This however can be justified by examining
the contributions of the remaining states, which necessarily
contain a larger number of strings. As in the case of $N=1$,
these lead to a larger number of Airy functions, hence
to subdominant asymptotics. 

\paragraph{Conclusion. ---}
We analyzed a general correspondence 
between random variables arising in very different contexts of statistical mechanics: 
on the one hand, the sum of the $\num$ largest eigenvalues in the GUE and on the other,
the free energy of $\num$ non-crossing directed polymers in a $d=1+1$ random media. 
We provided a striking indication that these two quantities have the same distributions 
for any $\num$, by comparing the tails of their PDF's at large positive values.
Indeed, the perfect agreement found between the Laplace transforms associated to the leading stretched 
exponential decays implies the non-trivial matching of an infinite series of coefficients. 
This naturally extends the 
well-known $\num=1$ case, where the single-polymer free energy, in turn the KPZ height,
maps to the largest eigenvalue of a GUE random matrix. In view of existing results
for DP discrete zero temperature models, it also nicely confirms universality for $N>1$.
Along the same line of ideas, one can put forward a more general conjecture,
where the joint distributions of the ensemble of non-crossing free energies in the same 
random medium is mapped into the joint distribution of the $N$ largest eigenvalues, i.e.
\bea
\frac{1}{t^{1/3}} \left\{\ln \frac{{\cal Z}_{1}}{{\cal Z}_{0}}, \ldots,\ln \frac{{\cal Z}_{\num}}{{\cal Z}_{\num-1}} \right\} \stackrel{\mbox{\tiny in law}}{ \equiv}  \{\gamma_1, \ldots, \gamma_\num\} \; . 
\eea 
It seems natural \cite{OConnellWarrenMultiLayer,CorwinKPZlineensemble}
as both ensembles of random variables involve strong correlations
which reflects in the distributions of the marginals (i.e. the partial sums) studied in this letter. 
A proof of this conjecture 
would be beneficial for a full understanding of the ubiquitous appearance of random matrix extreme 
statistics. 

\paragraph{Acknowledgements. ---}
We thank A. Borodin, I. Corwin for careful reading of the manuscript
and useful comments. We are also grateful to P. Di Francesco, J. Quastel, 
N. O'Connell, G. Schehr, S. Majumdar
and J. Warren
for discussions.
This work is supported by ``Investissements d'Avenir'' 
LabEx PALM (ANR-10-LABX-0039-PALM)
and by PSL grant ANR-10-IDEX-0001-02-PSL.





\onecolumngrid
\newpage 

\appendix
\setcounter{equation}{0}
\renewcommand{\thetable}{S\arabic{table}}
\renewcommand{\theequation}{S\arabic{equation}}
\renewcommand{\thefigure}{S\arabic{figure}}

\begin{center}
{\Large Supplementary Material for EPAPS \\ 
\titleinfo
}
\end{center}
Here we give additional details about the calculations presented in the letter.

\section{Tail of the partial sums of GUE edge eigenvalues}

Let us start from the definition in Eq.~\eqref{lapltrans}, where $\rhotail_\num(\gamma)$ is obtained 
by \eqref{sumdistr}, replacing $\PDGUE_\num(\gv_\num) \to r_\num(\gv_\num)$. 
\begin{multline}
N! ~ \trhoGUE_N  (u) = \prod_i \int_{-\infty}^\infty d\gamma_i e^{\gamma_i u} 
\det [\KAi(\gamma_i, \gamma_j)]_{i,j=1}^\num =
    \sum_{P \in \mathcal{S}_\num} (-1)^P \int_{\gamma_i \in \mathbb{R}} \int_{v_i >0} 
  \prod_j e^{\gamma_j u} \Ai(\gamma_j + v_j) \Ai(\gamma_{P_j} + v_j)  = \\=
   \sum_{P \in \mathcal{S}_\num} (-1)^P \int_{\gamma_i \in \mathbb{R}} \int_{v_i >0} 
  \prod_j e^{\gamma_j u} \Ai(\gamma_j + v_j) \Ai(\gamma_{j} + v_{P_j})  
 =  \int_{v_i > 0} 
 \det \big[ \int_{-\infty}^\infty d\gamma e^{\gamma u} \Ai(\gamma + v_j) \Ai(\gamma + v_k) \big]_{j,k=1}^\num 
\end{multline}
which is Eq.~\eqref{hatp} in the main text. Using the identity
\begin{equation}
 \int_{-\infty}^\infty dq \Ai(q^2 + v + v') e^{i q(v-v')} = 2^{2/3} \pi \Ai(2^{1/3} v) \Ai(2^{1/3}v')
\end{equation}
the integral over $\gamma$ can be performed thanks to the 
integral formula in Eq.~\eqref{airytrick}. Finally, the integral over $k$ reduces to a gaussian integral and leads to 
\be \label{mainsuppl}
\trhoGUE_N  (u) = \frac{e^{\frac{\num u^3}{12}}u^{-\frac{3N}{2}}}{\pi^{\num/2} \num!}
 \prod_{i=1}^\num \int_{v_i>0} \; e^{-2 v_i} \det \left[
 e^{-\frac{(v_j-v_k)^2}{u^3}} \right]_{j,k=1}^\num
\ee
after rescaling $v_j \to 2 v_j/u$, which is Eq.~\eqref{main} in the main text.

We now study the asymptotic behavior at large $u$. Since the determinant is symmetric and vanishes whenever $v_i = v_j$, we have at the leading order in $1/u$
\begin{equation}
\label{uexp}
\det \left[
 e^{-\frac{(v_i-v_j)^2}{u^3}} \right]_{i,j=1}^\num \simeq \frac{1}{G(\num+1)}  
\prod_{i<j} \frac{2(v_i - v_j)^2}{u^3} \;.
\end{equation}
In this expression, the prefactor can be fixed by setting $v_i = i$ and computing explicitly the left-hand side
\begin{equation}
e^{- \frac{2}{u^3}\sum_i i^2} \det \left[
 e^{\frac{2 i j}{ u^3}} \right]_{i,j=1}^N = e^{- \frac{2}{u^3}\sum_i (i^2 + i)} \prod_{i<j} ( e^{\frac{2 j}{u^3}} -
 e^{\frac{2 i}{u^3}}) \simeq 2^{N(N-1)/2} u^{-3N(N-1)/2} \prod_{i<j} (j-i) \;.
 \end{equation}
Then, by comparing with the right-hand side and using that $\prod_{i<j} (j-i) = G(N+1)$,
we arrive at Eq.~\eqref{uexp}. 
Note that (\ref{uexp}) is a particular case of the more general identity for any function $f(x,y)$, differentiable near zero
\begin{equation}
 \det \left( f(\epsilon x_i, \epsilon x_j) \right)_{i,j=1}^N  = \epsilon^{N(N-1)} \frac{\prod_{i<j} (x_i - x_j)^2}{G(N+1)^2}
 \left.\det \left[ \partial_x^{i-1} \partial_y^{j-1} f(x, y) \right]_{i,j=1}^N \right|_{x=y=0} \;.
\end{equation}
valid to leading order in small $\epsilon$. Indeed, to that order, it is equivalent to insert $f(x,y)=e^{2 x y/u^3}$ 
whose determinant of derivatives is simply $G(N+1)$. This equation has been used in the
text to arrive at (\ref{det1}). 

The integral in \eqref{main} can be computed again using the Selberg integral. In particular we have
\begin{equation}
\label{wishartselberg}
  \prod_{i=1}^n \int_0^\infty dz_i z_i^{\alpha-1} e^{-k z_i} \prod_{1 \leq i<j \leq n} |z_i - z_j|^{2\gamma} = 
\prod_{j=0}^{n-1} \frac{\Gamma(\alpha + j \gamma) \Gamma(1 + (j+1) \gamma)}{k^{\alpha + (n-1)\gamma}\Gamma(1 + \gamma)}
\end{equation}
which for $\gamma = \alpha = 1$ and $k=2$, simply reduces to $G(N+1) G(N+2) 2^{-N^2}$. When inserted in \eqref{main}, it leads to
Eq.~\eqref{pNtilde1} in the main text.

\section{Leading contribution to the $m$-th moment: term with the smallest number of strings, 
$n_s=N$.}

Let us find the leading non-vanishing contribution $\mathcal{B}_{N,\mom}[\kv, \mv]$.
The first non-vanishing contribution comes from the configuration composed by 
$N$ $\mom$-strings, i.e. $\partNS = \num$ and $m_j = \mom$. 
We come now to the problem of evaluating $\mathcal{B}_{N,\mom}(\pv)$ for this particular configuration. 
From its definition, one can show that it is a symmetric polynomial \cite{UsNonCrossing2}
in the variables $k_j$, hence of degree $N(N-1) m$. It is strongly constrained by
the fact that $\mathcal{B}_{\num,\mom}(\pv)=0$ whenever $\pv$ contains a sequence
$k,k+i,..k+i m$ for some $k$. This implies that $\mathcal{B}_{N,\mom}(\kv) =0$ when
$k_i = k_j + i p$ with $p=1,..m$. This implies that 
\begin{equation} \label{Bres} 
\mathcal{B}_{N,\mom}(\kv) = \frac{\mom!^N}{(\mom N)!}\prod_{1 \leq i<j \leq \num} \prod_{p=1}^\mom[(k_i - k_j)^2 + p^2 ]
\end{equation}
where the normalization is fixed by the limit where all $k_j$ are large, in which case
the evaluation of (\ref{Bdef}) is a simple exercise in combinatorics. The energy and normalization factor for this configuration have the form
\begin{equation}
 E(\kv, \mv) = \mom \sum_{j=1}^N k_j^2 + \frac{1}{12} (\mom - \mom^3)
 \quad , \quad \Phi(\kv, \mom) = 
 \prod_{1\leq j<j'\leq \num} 
\frac{(k_j - k_{j'})^2}{(k_j - k_{j'})^2 +  \mom^2}\;.
\end{equation}
Multiplying the second equation with (\ref{Bres}) one finds equation (\ref{resB2}) in the text
with $(x)_m=x(x+1)..(x+m-1) = \Gamma(x+m)/\Gamma(x)$. 
Inserting
in \ref{partsum} keeping only the configuration $m_j=m$, $j=1,..n_s$ with $n_s=N$ one finds
 \begin{align}
\label{Zres}
\mathcal{Z}^{(0)}_{N,m}(t) :=  \overline{ \hat{\mathcal{Z}}_{\num} (t)^\mom}|_{m_j=m} = 
\frac{ \mom!^\num e^{-\frac{\num t}{12} (\mom - \mom^3)}}{\num! \mom^\num (2 \pi)^{\num}} 
 \prod_{j=1}^{\num} \int_{-\infty}^{+\infty} dk_j e^{-m k_j^2 t}
\prod_{1 \leq i<j \leq \num} \prod_{p=0}^{\mom-1}[(k_i - k_j)^2 + p^2 ] \;.
\end{align}
Note that this is {\it the exact contribution to the $m$-th moment for all $t$} of the $N$ $m$-string state,
which is also {\it the state with the lowest number of strings $n_s=N$}.

What is the significance of the contribution of this state? before discussing that point, let us
recall the analysis for $N=1$. Then (\ref{Zres}) reduces to 
\bea
\mathcal{Z}^{(0)}_{1,m}(t) = 
\frac{ \mom! e^{-\frac{\num t}{12} (\mom - \mom^3)}}{ \mom (2 \pi)} 
\int_{-\infty}^{+\infty} dk e^{-m k^2 t} = 
\frac{ \mom! e^{-\frac{\num t}{12} (\mom - \mom^3)}}{m^{3/2} \sqrt{4 \pi t}} \label{1string} 
\eea 
which is the well known single string contribution $n_s=1$ associated to
the droplet initial condition. Note that it is valid for all times $t$. 
It contains information about the right tail of the PDF, $P(H,t)$, of the
KPZ height, where we denote $H=h(0,t)+ \frac{t}{12}$. More precisely it 
contains two types of information (i) the right tail of the TW distribution for typical fluctuations, i.e. large values of
$H$ within the regime $H \sim t^{1/3}$, (ii) the form of $P(H,t)$ in the 
large deviation regime, i.e. for atypically large fluctuations $H \sim t$. In the
large deviation regime, from (\ref{1string}) one finds (see the Supp Mat in Ref. \cite{largedev}) that
at large $t$ 
\bea \label{subdom} 
\ln P(H,t) \simeq { - t \frac{4}{3} z^{3/2} -  \ln t  - \chi(z) + o(1) } \quad , \quad z=H/t \quad \text{fixed} \;.
\eea 
where $\chi(z)$ encodes the first subleading correction \cite{largedev}
\bea \label{chiinteger} 
\chi(z) = \ln( 4 \pi)  + \frac{1}{2} \ln z  - \ln\left(\Gamma(2 \sqrt{z})\right) 
\eea
Note that the integer moments themselves a-priori allow to determine $\chi(z)$ 
only for $z=z_m=m^2/4$ for $m \in \mathbb{N}^*$. However the resummation
of these contributions into a generating function, i.e. $g^{(0)}(s)$ as performed here in the text,
allows to obtain $\chi(z)$ for all $z$ (see the derivation in \cite{largedev}). 
In the limit $z \to 0$ using that
$\Gamma(2 \sqrt{z}) \simeq 1/(2 \sqrt{z})$ this expression matches with the right tail of
the TW distribution, i.e. $P(H,t)  \simeq \frac{1}{t^{1/3}} f_2\left(\frac{H}{t^{1/3}}\right)$ with $
f_2(x) \simeq \frac{1}{8 \pi x} e^{- \frac{4}{3} x^{3/2}}$ for $x \to +\infty$. 

Here, we extend this analysis to $N>1$. The first property of (\ref{Zres}) is that it gives the leading behavior of the $m$-th moment at fixed $m,N$ and large $t$, up
to an exponential correction, for $m \geq 2$, more precisely
\bea
 \overline{ \hat{\mathcal{Z}}_{\num} (t)^\mom} = 
 \mathcal{Z}^{(0)}_{N,m}(t)  (1 + O( e^{- \frac{1}{4} m (m-1) t } ) ) 
\eea 
where the subdominant terms come from lowest "excitations" with a larger number of
strings, i.e. such that $n_s=N+1$
(with $m_j=m$, $j=1,..N-1$, $m_N=m-1$, $m_{N+1}=1$).

The dominant term in the limit of large $t$ in Eq. (\ref{Zres}) has the form
\begin{equation}
\label{Zres2}
 \overline{ \mathcal{Z}_{\num} (t)^\mom } 
 \simeq 
\frac{ (\mom-1)!^{\num^2} e^{-\frac{\num t}{12} (\mom - \mom^3)} G(N+1)}
{(\sqrt{2 \pi})^{\num} (2mt)^{N^2/2}} .
\end{equation}
Indeed, since $k_j$ scales as $1/\sqrt{t}$, this term can be obtained by replacing in the product $\prod_{p=0}^{\mom-1}[(k_i - k_j)^2 + p^2 ] \to 
(k_i - k_j)^2  \prod_{p \geq 1}^{\mom-1} p^2$. The remaining integral over the $k_j$ can then be performed
using the Mehta integral $\int_{-\infty}^{+\infty} dp_j e^{-p_j^2/2} 
\prod_{1 \leq i<j \leq \num} (p_i - p_j)^2 = (2 \pi)^{N/2} G(N+2)$. 

From a saddle point argument similar to the one given in \cite{largedev}, one finds from the formula \eqref{Zres2} that the PDF, $P_N(H,t)$ of the variable 
$H= N t/12 + \ln \hat{\mathcal{Z}}_{\num} (t)$, should take the form
\bea \label{subdomN} 
\ln P_N(H,t) \simeq { - t \frac{4}{3 \sqrt{N}} z^{3/2} - a_N \ln t  - \chi_N(z) + o(1) } \quad , \quad z=H/t \quad \text{fixed} \;.
\eea 
in the large deviation regime. Indeed 
\bea
\overline{e^{m H} } \simeq t \int dz e^{- t (\frac{4}{3 \sqrt{N}} z^{3/2} - m z) - a_N \ln t  - \chi_N(z) }
\simeq t^{\frac{1}{2}-a_N} \sqrt{\pi m N} e^{ N t \frac{m^3}{12} - \chi_N(z_m)} 
\eea 
where the saddle point is located at $z_m = N m^2/4$. Matching with the
formula \eqref{Zres2} we find
\bea
&& a_N = \frac{1+N^2}{2} \\
&& \chi_N(z) = - N^2 \ln \Gamma(2 \sqrt{\frac{z}{N}}) 
+ \frac{N^2}{2} \ln (4 \sqrt{\frac{z}{N}}) + \frac{1}{2} \ln(2 \pi N \sqrt{\frac{z}{N}}) 
+ \frac{N}{2} \ln(2 \pi) - \ln G(N+1) 
\eea 
which is valid a priori for $z=z_m$ for $m \in \mathbb{N}^*$. One could conjecture that it remains the correct 
analytic continuation for arbitrary $z>0$. To confirm or infirm this conjecture
one would need to analyze the formula (\ref{s2}) for the generating function
$g^{(0)}(s)$ in the regime where $s \sim t^{2/3}$, as was done for $N=1$
in \cite{largedev}. We leave this for future work. 

In the limit $z \to 0$ one should be able to match with the tail of the
typical values for $H \sim t^{1/3}$ which has been calculated in this paper.
Such a check can be performed on formula (\ref{Zres2})
directly. Assuming the simplest analytic continuation $(m-1)! \to \Gamma(m)$
in that formula one can check explicitly that the Laplace
transform $\tilde \rho_N(u)$ obtained in (\ref{pNtilde1}) when evaluated at the argument $u= m t^{1/3}$ 
recovers exactly the small $m$ limit of (\ref{Zres2}).

\section{Summation of the generating function}
We now perform the explicit derivation of Eq.~\eqref{gnexpr} given in the text. We insert 
\eqref{Zres} in 
\begin{multline}
g^0_\num(s) = \sum_{\mom = 0}^\infty \frac{(-1)^\mom}{m!} x^m
\mathcal{Z}^{(0)}_{N,m}(t)=
\prod_{j=1}^{\num} \int_{-\infty}^{+\infty} \frac{dk_j}{2\pi} \sum_{\mom = 0}^\infty \frac{(-1)^\mom}{m!}  e^{-\lambda s m}
\frac{ \mom!^\num e^{\frac{\num \mom^3 t}{12} }}{\num! \mom^\num} 
 e^{-m k_j^2 t}
\prod_{1 \leq i<j \leq \num} \frac{\Gamma(m - \imath k_{ij}) \Gamma(m + \imath k_{ij}) }{\Gamma(- \imath k_{ij}) \Gamma(\imath k_{ij})}=\\
=
\prod_{j=1}^{\num} \int_{-\infty}^\infty dy \Ai(y) \int_{-\infty}^{+\infty} \frac{dk_j}{2\pi} \sum_{\mom = 0}^\infty  \frac{(-1)^\mom}{m!} e^{\lambda (y-s) m}
\frac{ \mom!^\num}{\num! \mom^\num} 
 e^{-m k_j^2 t}
\prod_{1 \leq i<j \leq \num} \frac{\Gamma(m - \imath k_{ij}) \Gamma(m + \imath k_{ij}) }{\Gamma(- \imath k_{ij}) \Gamma(\imath k_{ij})}
\end{multline}
where in the second line we used the Airy-trick introduced in Eq.~\eqref{airytrick} in the main text, with 
$w = m (\num t /4)^{1/3} = m \lambda$. 
We now use the Mellin-Barnes formula to perform the summation over $m$ in the
large time limit as explained in Eq.~\eqref{MBtrick}. This amounts to replacing the integer variable
$m$ with the integral over the complex $z$, i.e.
\begin{equation} \label{s2} 
g^0_\num(s) = - \prod_{j=1}^{\num} \int_{-\infty}^\infty dy \Ai(y) \int_{-\infty}^{+\infty} \frac{dk_j}{2\pi} 
 \int_{\epsilon - \imath \infty}^{\epsilon + \imath \infty} \frac{dz}{2\imath \sin(\pi z)} 
e^{\lambda (y-s) z}
\frac{ \Gamma(z+1)^{\num-1}}{\num! z^\num} 
 e^{-z k_j^2 t}
 \prod_{1 \leq i<j \leq \num} \frac{\Gamma(z - \imath k_{ij}) \Gamma(z + \imath k_{ij}) }{\Gamma(- \imath k_{ij}) \Gamma(\imath k_{ij})}\;.
\end{equation}
We now perform the changes of variables in the integrals $z \to z/\lambda, k_{i} \to k_{i} \to k_{i} \sqrt{\num}/(2\lambda)$ and then 
$ y \to y + s + \sum_i k_i^2$ and take
the large time limit: $\lambda\to\infty$:
\begin{multline} 
g^0_\num(s) = -\frac{\num^{\num/2}}{2^{\num}\num! }
\prod_{j=1}^{\num} \int_{-\infty}^{+\infty} \frac{dk_j}{2\pi} \int_{-\infty}^\infty dy \Ai(y + s + \sum_i k_i^2)  
\int_{\epsilon - i \infty}^{\epsilon + i \infty} dz \; \frac{ \Gamma(\frac{z}{\lambda}+1)^{\num-1} z^{-\num} e^{y z}
}{2 \imath \lambda \sin(\pi z/\lambda)} 
 \prod_{1 \leq i<j \leq \num} \frac{\Gamma\bigl(\frac{2 z - \imath \sqrt{\num} k_{ij}}{2 \lambda}\bigr) \Gamma\bigl(\frac{2 z + \imath \sqrt{\num} k_{ij}}{2 \lambda}\bigr) }
{\Gamma\bigl(- \frac{\imath \sqrt{\num} k_{ij}}{2 \lambda}\bigr) \Gamma\bigl(\frac{\imath \sqrt{\num} k_{ij}}{2 \lambda}\bigr)} \\ \stackrel{\lambda \to \infty}{=} 
-\frac{\num^{\num/2}}{2^{\num}\num! }
\prod_{j=1}^{\num} \int_{-\infty}^{+\infty} \frac{dk_j}{2\pi} \int_{-\infty}^\infty dy  \Ai(y + s + \sum_i k_i^2) \int_{-\infty}^{+\infty} 
\int_{\epsilon - i \infty}^{\epsilon + i \infty} \frac{dz \, e^{y z}}{2\pi \imath z^{\num + 1}}  
\; 
\prod_{1 \leq i<j \leq \num} \frac{ (k_{i} - k_j)^2 \num}
{4 z^2 +  (k_{i} - k_j)^2 \num} 
\end{multline}
Finally we use the following identity:
\bea
\label{dettransf}
 \prod_{1 \leq i<j \leq \num} \frac{(k_i-k_j)^2 N}{4z^2 + (k_i-k_j)^2 N} = 
{\rm det}[ \frac{1}{i (k_i-k_j) \sqrt{N} + 2 z}]_{i,j=1}^N (2 z)^N 
\eea 
and then we replace $z \to \sqrt{\num} z$ to obtain the Eq.~\eqref{gnexpr} in the main text.

\section{Laplace transform of the DP free energy distribution}
We derived in the text the expression
\begin{equation}
\label{gnexpr1}
 g^{(0)}_N(s) \stackrel{t \to \infty}{=}
 1- \frac{1}{\num!}\prod_{i=1}^{\num} \int_{-\infty}^{+\infty} \frac{dk_i}{2\pi} 
   \int_{0}^\infty dy \Ai\bigl(y+\sum_i k_i^2 + s\bigr) 
   \int_{\epsilon - \imath \infty}^{\epsilon + \imath \infty} \frac{dz}{2\pi \imath z} 
e^{\sqrt{N} z y} \det \left[\frac{1}{2z + \imath k_{jk}}\right]_{j,k=1}^N \;.
\end{equation}
The Laplace transform defined in
in \eqref{lapltransDP}, can then be performed using that:
\begin{equation}
\label{aitrickder}
  \int_{-\infty}^\infty ds \Ai'(y+ \sum_i k_i^2+s) e^{(\frac{N}{4})^{1/3} u s} =  
   - \tilde u e^{-\tilde u (\sum_i k_i^2 + y)} 
  e^{\frac{N u^3}{12}}\;, \qquad \tilde u \equiv (N/4)^{1/3} u\;.
\end{equation}
Then, substituting Eq.~\eqref{gnexpr1} in Eq.~\eqref{lapltransDP}, and using Eq.~\eqref{aitrickder} to perform the integral
over $s$, we have, after integrating over $y$:
\begin{multline}
\trhoDP_\num(u) = 
\frac{\tilde u e^{\frac{N u^3}{12}}}{\num!} \prod_{i=1}^{\num} \int_{-\infty}^{+\infty} \frac{dk_i \; e^{-\tilde u k_i^2}}{2\pi}  
    \int_{\epsilon - \imath \infty}^{\epsilon + \imath \infty} \frac{dz}{2\pi \imath z} 
\frac{1}{\tilde u - \sqrt{N}z}\det \left[\frac{1}{2z + \imath k_{jk}}\right]_{j,k=1}^N =
\\ =
\frac{e^{\frac{N u^3}{12}}}{\num!} \prod_{i=1}^{\num} \int_{-\infty}^{+\infty} \frac{dk_i \; e^{-\tilde u k_i^2}}{2\pi} 
\sum_{P\in \mathcal{S}_\num} (-1)^P \prod_{j=1}^\num \frac{1}{2\tilde uN^{-1/2} + \imath k_{j P_j}}=\\=
\frac{e^{\frac{N u^3}{12}}}{\num!} \sum_{P\in \mathcal{S}_\num} (-1)^P \prod_{i=1}^{\num} \int_{-\infty}^{+\infty} \frac{dk_i \; e^{-\tilde u k_i^2}}{2\pi} 
\prod_{j=1}^\num \int_0^\infty dv_j e^{-2\tilde uN^{-1/2}v_j} + e^{-\imath( k_{j} - k_{ P_j})v_j}=\\=
\frac{e^{\frac{N u^3}{12}}}{\num!} \sum_{P\in \mathcal{S}_\num} (-1)^P \prod_{i=1}^{\num} \int_{-\infty}^{+\infty} \frac{dk_i \; e^{-\tilde u k_i^2}}{2\pi} 
\prod_{j=1}^\num \int_0^\infty dv_j e^{-2\tilde uN^{-1/2}v_j} + e^{-\imath k_{j} (v_j - v_{P_j})}=
\\= \frac{e^{\frac{N u^3}{12}}}{2^{N} \pi^{N/2}\num!}\prod_{i=1}^{\num} \int_{v_i>0} dv_i e^{-v_i u^{3/2}}    
{\rm det}[ e^{-\frac{(v_j-v_k)^2}{4}}]_{j,k=1}^\num \label{last}
\end{multline}
where in the second line we integrated over $z$, by closing the contour at $\Re[z] > 0$ and in the fourth line
we redefined the permutation as $P \to P^{-1}$. To obtain the last line we rescaled $v_j \to \sqrt{\tilde u} v_j$.

Now, upon the change of variable $v_i \to 2 v_i/u^{3/2}$ one puts (\ref{last}) in exactly the same
form as (\ref{main}) in the text, hence showing that $\trhoGUE_\num (u) = \trhoDP_\num (u)$. 


\begin{thebibliography}{100}
\bibitem{footnote0}
The $-N t/12$ in (\ref{larget}) comes from the fact that one normalizes $\hat {\cal Z}_\num(t)$
using Ito rule which implies $\overline{\hat {\cal Z}_\num(t)}=\hat {\cal Z}^{\eta=0}_\num(t)$,
i.e. so that the average coincides with the partition sum in the absence of disorder.
\end{thebibliography}

\end{document}